\documentclass[prd,twocolumn]{revtex4}
\usepackage{natbib}
\usepackage{epsfig}
\usepackage{bm}
\usepackage{color}

\providecommand{\eprint}[1]{\href{http://arxiv.org/abs/#1}{#1}}

\providecommand{\adsurl}[1]{\href{#1}{}}

\begin{document}

\newcommand{\clee}{C_{\ell}^{EE}}
\newcommand{\clbb}{C_{\ell}^{BB}}
\newcommand{\cltens}{C_{\ell}^{BB,{\rm T}}}
\newcommand{\cllens}{C_{\ell}^{BB,{\rm L}}}
\newcommand{\xef}{x_e^{\rm fid}}
\newcommand{\xet}{x_e^{\rm true}}
\newcommand{\dz}{\Delta z}
\newcommand{\zmax}{z_{\rm max}}
\newcommand{\zmin}{z_{\rm min}}
\newcommand{\zmid}{z_{\rm mid}}
\newcommand{\lmax}{\ell_{\rm max}}
\newcommand{\lcdm}{$\Lambda$CDM}
\newcommand{\wmap}{\emph{WMAP}}
\newcommand{\planck}{\emph{Planck}}
\newcommand{\cmbpol}{\emph{CMBPol}}

\newcommand{\mnras}{Mon. Not. R. Astron. Soc.}
\newcommand{\aap}{Astron. Astrophys.}
\newcommand{\apjs}{Astrophys. J. Suppl. Ser.}

\title{Impact of reionization on CMB polarization tests of slow-roll inflation}
%\shorttitle{}

\author{Michael J. Mortonson$^{1,2}$ and Wayne Hu$^{1,3}$}
\affiliation{$^{1}$Kavli Institute for Cosmological Physics, 
Enrico Fermi Institute, University of Chicago, Chicago, IL 60637\\
$^{2}$Department of Physics,  University of Chicago, Chicago, IL 60637\\
$^{3}$Department of Astronomy and Astrophysics, University of Chicago, Chicago, IL 60637
}

\date{\today}

\begin{abstract}
Estimates of inflationary parameters from the
CMB $B$-mode polarization spectrum on the largest scales depend on knowledge of
 the reionization history, especially at low
tensor-to-scalar ratio. 
Assuming an incorrect reionization history in the analysis of 
such polarization data can strongly bias the inflationary parameters. 
One consequence is that the single-field slow-roll consistency relation 
between the tensor-to-scalar 
ratio and tensor tilt might be excluded with high significance even if 
this relation holds in reality. 
We explain the origin of the bias and present case studies 
with various tensor amplitudes and noise characteristics. 
A more model-independent approach can account for uncertainties
 about reionization, and we show that parametrizing 
the reionization history by a set of its principal components with respect to 
$E$-mode polarization removes the bias in inflationary parameter measurement with little
degradation in precision.
\end{abstract}

%\keywords{cosmic microwave background --- cosmology: theory --- large-scale structure of universe}

\maketitle

% =====================================================
\section{Introduction}
\label{sec:intro}

Temperature and polarization power spectra of the cosmic microwave background
(CMB) are consistent with predictions
of the simplest inflationary models~\cite{Gut81,AlbSte82,Lin82,Sat81}:
a nearly flat geometry,
superhorizon correlations as probed by the spectrum of acoustic peaks, 
and primordial scalar perturbations that are adiabatic, Gaussian, and close to 
scale-invariant~\cite{HuWhi96c,SpeZal97,HuSpeWhi97,Peietal03,Speetal07}.
One of the key remaining signatures of inflation, tensor 
perturbations (i.e.\ gravitational waves)~\cite{KamKosSte97,SelZal97}, 
has yet to be detected. 
Depending on the amplitude of 
the tensor perturbations, which is not well constrained theoretically,
it may be possible to measure the angular power spectrum of the inflationary 
gravitational waves in the $B$-mode component of the CMB polarization 
on large scales.
Non-detection of the tensor spectrum does not necessarily rule out 
inflation, but upper limits on $r$ can be used to exclude particular 
models of inflation and limit its energy scale.
Many experiments have been proposed to search for this 
signal~\cite{Planck,CMBTaskForce,Oxletal04,LawGaiSei04,Ruhl:2004kv,Mafetal05,Yooetal06,Kogut:2006nd,Macetal07,Polenta07}.

Measurement of  tensor perturbations in the $B$-mode 
polarization power spectrum would 
test models of inflation by constraining inflationary parameters.
These parameters include the tensor-to-scalar ratio, $r$, and the 
tensor spectral index, $n_t$, which are related by a consistency relation 
under the simplest single-field slow-roll inflationary scenarios.
If the tensor spectrum can be detected, precise measurements over a wide 
range of scales could test the consistency relation.

CMB constraints on $r$ and $n_t$ depend on the ability to 
accurately determine the large-scale power in $B$-modes due to tensor perturbations, 
independent of the effects of other cosmological parameters. 
On the largest scales, the tensor $B$-mode spectrum
 depends not only on inflationary parameters 
but also on the reionization history of the universe~\cite{Zal97}. 
The main impact of reionization on the spectrum is through the total optical 
depth, $\tau$.  The 3-year \wmap\ measurements of $E$-mode polarization 
determine $\tau$ to an accuracy of about $30\%$~\cite{Pagetal07,Speetal07}, 
and future CMB experiments should constrain $\tau$ at the~$5-10\%$~level~\cite{Holetal03,KeaMil06,Planck,MorHu07b}.

However, the detailed evolution of the reionization history also significantly 
affects the large-scale polarization spectra.  Uncertainty in this history leads
to added uncertainty in inflationary 
parameters.  Moreover, incorrect inferences due to an
oversimplified treatment of reionization may bias estimates of 
inflationary parameters. 
For unbiased estimation of the optical depth from the $E$-mode reionization peak, 
the solution is to use a complete, principal-component-based 
description of reionization effects  when 
estimating parameters from CMB polarization data~\cite{HuHol03,MorHu07b}.
In this paper, we extend this approach to tensor $B$-mode polarization
and show that it is equally if not more effective in ensuring accurate measurements
with little loss in precision.

The outline of the paper is as follows.
We discuss the effects of reionization and inflationary parameters on 
the polarization power spectra and the large-scale degeneracy between 
these parameters in \S~\ref{sec:param}. A brief overview of the principal 
component parametrization of the reionization history follows in \S~\ref{sec:pcs}. 
In \S~\ref{sec:mcmc}, we describe our Markov Chain Monte Carlo analysis of 
simulated polarization data and give the resulting constraints on $\tau$, 
$r$, and $n_t$, which we discuss further in \S~\ref{sec:discuss}.

% ****************************************
\begin{figure}
\centerline{\psfig{file=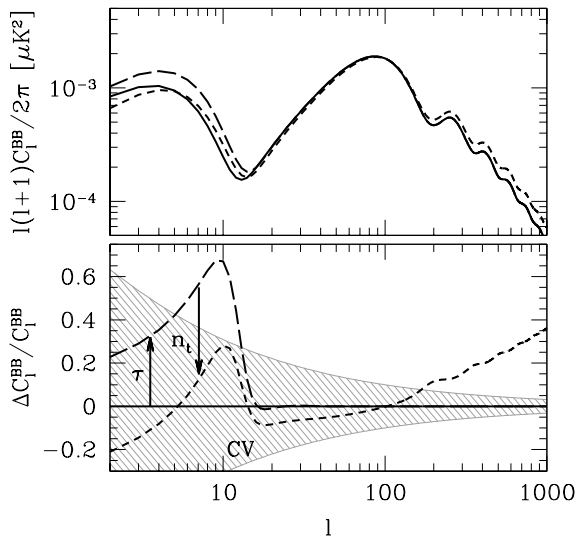, width=3.0in}}
\caption{$B$-mode tensor spectra 
 illustrating the degeneracy 
between $\tau$ and $n_t$ for large-scale measurements, 
with angular power spectra plotted in the upper panel and fractional 
deviations from the base model in the lower panel. 
For the base model (\emph{solid}), $r=0.03$, $\tau=0.1$, and 
$n_t=-0.00375$. The other two models have 
$\{\tau,n_t\}=\{0.12,-0.00375\}$ (\emph{long dashed}) and 
$\{0.12,0.13\}$ (\emph{short dashed}), with a pivot scale of 
$k_{\rm pivot}=0.01~{\rm Mpc}^{-1}$.
The reionization history here is assumed to be instantaneous. 
Cosmic variance of $\cltens$ for the base model, which excludes the variance
from lensing, is shown by the shaded band in 
the lower panel.
}
\label{fig:degen}
\end{figure}
% ****************************************

% =====================================================
\section{Polarization dependence on reionization and inflationary parameters}
\label{sec:param}

Like the scalar $E$-mode polarization power spectrum, the tensor $B$-mode spectrum
$\cltens$ consists of 
two main components: one from recombination that peaks at $\ell \sim 100$, 
and the other from the epoch of reionization, 
which peaks near $\ell \sim 5$ and dominates at $\ell \lesssim 20$ (Fig.~\ref{fig:degen}).   
In the tensor power spectrum, these components arise from wavenumbers
$k \sim 0.01$ Mpc$^{-1}$  and $\sim 0.0007$ Mpc$^{-1}$ respectively.   They provide a lever arm of 
over a decade in physical scale for measuring the tensor tilt. This
lever arm is slightly shortened compared with the ratio of angular scales due
to the closer distance to reionization.   Nonetheless, these features provide an opportunity
to test the slow-roll inflationary consistency relation~\cite[e.g.,][]{Lidetal97},
\begin{equation}
n_t = -r/8,
\label{eq:consrel}
\end{equation}
 between the tensor-to-scalar ratio $r$
and tensor tilt $n_t$. Deviations from the consistency relation and
running of the tilt come in at second order in the slow-roll parameters.  Specifically, $r$ is four times the ratio of the 
tensor power spectrum amplitude (for one component of gravitational waves) 
to the scalar curvature power spectrum amplitude, in accordance with the 
definition used by CAMB and \wmap.

Note that we quote $r$ at a normalization scale of $k_{\rm pivot} = 0.01$ Mpc$^{-1}$
whereas 3-year \wmap\ results quote it at $k_{\rm pivot}=0.002$ Mpc$^{-1}$ and CAMB defaults 
to $k_{\rm pivot}=0.05$ Mpc$^{-1}$.  Our choice, corresponding to the recombination peak at 
$\ell \approx 100$, better reflects the scale at which the tensor spectrum can be best measured.
We will use subscripts to denote choices of scale other than $0.01$~Mpc$^{-1}$, 
e.g. $r_{0.05}$ for normalization at $k_{\rm pivot}=0.05$~Mpc$^{-1}$, so that
\begin{equation}
r_{k_{\rm pivot}/{\rm Mpc}^{-1}} = r \left(\frac{k_{\rm pivot}}{0.01~{\rm Mpc}^{-1}}\right)^{n_t+1-n_s},
\label{eq:rscale}
\end{equation}
assuming no running of the scalar or tensor spectral indices.
Since we only consider small deviations from scale-invariance, the tensor-to-scalar 
ratio varies little with the normalization scale so any corrections to 
Eq.~(\ref{eq:consrel}) due to changing $k_{\rm pivot}$ are correspondingly small.

Cosmic variance and reionization history uncertainties in the interpretation of the reionization peak
limit the ability to measure the spectrum from this technique.    
In Fig.~\ref{fig:degen}, we show that
variation  in the optical depth to reionization, $\tau$, can mimic changes to 
$n_t$ within the precision of cosmic variance of the individual $\ell$ modes,
\begin{equation}
{\Delta \cltens \over \cltens} \approx \sqrt{2 \over 2\ell +1}\,.
\end{equation}

\begin{figure}
\centerline{\psfig{file=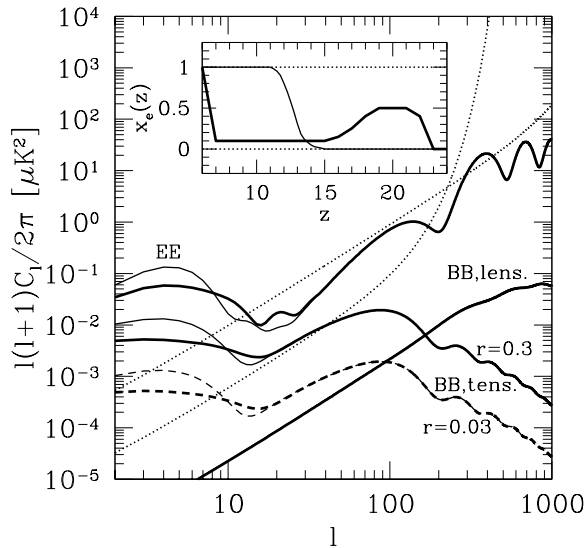, width=3.0in}}
\caption{$E$- and $B$-mode polarization angular power spectra and 
reionization histories (\emph{inset}) for an extended, 
``double'' reionization history with $\tau = 0.090$ (\emph{thick curves}) 
and a  nearly-instantaneous reionization model with optical depth 
$\tau = 0.105$ (\emph{thin}). 
For the $B$-modes, both the tensor spectra [with tensor-to-scalar 
ratio $r=0.3$ (\emph{solid}) and $r=0.03$ (\emph{dashed})] 
and lensing spectra are plotted. Dotted curves are the assumed 
\planck\ and \cmbpol\ noise power spectra, from top to bottom 
at low $\ell$ respectively.
%\vskip 0.25cm
}
\label{fig:cl}
\end{figure}

The example of Fig.~\ref{fig:degen} shows that there are at least two ways of breaking this 
degeneracy. One is by measuring the tensor spectrum at $\ell>100$, where 
the tilt of the spectrum matters but the optical depth has no effect 
other than an overall rescaling of the spectrum by $e^{-2\tau}$, which 
we absorb by changing the scalar amplitude $A_s$ to keep
$A_s e^{-2\tau}$ fixed.
Just as in the scalar case, if the tensor spectrum can be precisely
measured beyond the recombination peak, the global constraint on $n_t$ will not
be sensitive to reionization.  Unlike the scalar case, $\cltens$ falls sharply at higher
$\ell$ and becomes masked by $B$-modes 
generated by lensing of $E$-modes~\cite{ZalSel98} ($\cllens$; see Fig.~\ref{fig:cl}).  The power in the lensing 
$B$-modes is expected to be greatest at $\ell \sim 1000$, a smaller scale 
than the reionization and recombination peaks of the tensor spectrum. 
The relative amplitude of the tensor and lensing contributions to $\clbb$ depends on the 
tensor-to-scalar ratio, which the 3-year \wmap\ 
data restrict to $r\lesssim 0.3$~\cite{Speetal07}.

If $r$ is near the current upper limits,  the lensing spectrum can be statistically subtracted to a large extent.  Nevertheless, instrumental and foreground limitations
may still prevent the extraction of information from scales beyond the recombination peak for
next generation experiments such as {\it Planck} \cite{Bowetal04,Tucetal05,VerPeiJim06, AmbCooKap07}.
At much lower $r$, the best CMB constraints on $n_t$ will come from the combination
of the recombination and reionization peaks unless the lensing signal can be
subtracted directly from the polarization maps~\cite{HuOka01,KnoSon02,KesCooKam02,SelHir03,Marian:2007sr}.

The $\tau-n_t$ degeneracy is also broken through the constraint 
on $\tau$ from the $E$-mode reionization peak, which depends on
$\tau$ but not $n_t$ as long as $r$ is small enough that $\clee$ is 
dominated by scalar perturbations.  
To break the degeneracy in this way, it is important to have accurate constraints
on $\tau$ that do not depend on overly simplistic assumptions about the reionization 
history.  

The shape of the 
reionization peak depends on the history of the spatially-averaged 
ionized fraction, $x_e(z)$.  In Fig.~\ref{fig:cl}, we illustrate two models with
similar optical depths but different reionization histories.  
A model-dependent analysis of the polarization 
power spectrum that uses an incorrect form of $x_e(z)$ can result 
in significant bias in the total optical depth to 
reionization~\cite{Kapetal03,Holetal03,Coletal05,MorHu07b}.
To the extent that the constraint on $n_t$ 
relies on measurements of the reionization peak of $\cltens$, it may be biased as well in such an analysis.  We shall see that
the use of model-independent principal components of the reionization history 
protects against such biases at little cost to the precision of
the inflationary test.

% ****************************************
% ****************************************

% =====================================================
\section{Principal Component Parametrization of Reionization}
\label{sec:pcs}

Following~\cite{HuHol03,MorHu07b}, we parametrize the reionization history 
as a free function of redshift by decomposing $x_e(z)$ into its
principal components with respect to the $E$-mode polarization of the 
CMB:
\begin{equation}
x_e(z)=\xef(z)+\sum_{\mu}m_{\mu}S_{\mu}(z),
\label{eq:mmutoxe}
\end{equation}
where the principal components, $S_{\mu}(z)$, 
are the eigenfunctions of the Fisher matrix that describes 
the dependence of $\clee$ on $x_e(z)$, $m_{\mu}$ are the amplitudes of the 
principal components for a particular reionization history, and 
$\xef(z)$ is the fiducial model at which the Fisher matrix is computed. 
In practice, we construct the principal components assuming $E$-mode
information only by taking the tensor-to-scalar ratio $r=0$.  For allowed
values of $r$, the information on the ionization history from the tensors
is subdominant to the scalars. 
The inverses of the eigenvalues of the Fisher matrix give the 
estimated variances of the principal components, $\sigma_{\mu}^2$, which 
determine the ordering of the components by requiring that 
$\sigma_{\mu}^2 < \sigma_{\mu+1}^2$.
The main advantage of using principal components as a basis for $x_e(z)$ is 
that only a small number of the components are required to completely 
describe the effects of reionization on large-scale CMB polarization, 
so we obtain a very general parametrization of the reionization history at 
the expense of only a few additional parameters.

The principal components are defined over a limited range in redshift, 
$\zmin<z<\zmax$, with $x_e=0$ at $z>\zmax$ and $x_e=1$ at $z<\zmin$. 
We take $\zmin=6$, $\zmax=30$, and constant $\xef(z)=0.15$ here, 
and refer to~\cite{MorHu07b} for 
further discussion of the choices of these and other parameters 
related to the principal components.

% ****************************************
\begin{figure}
\centerline{\psfig{file=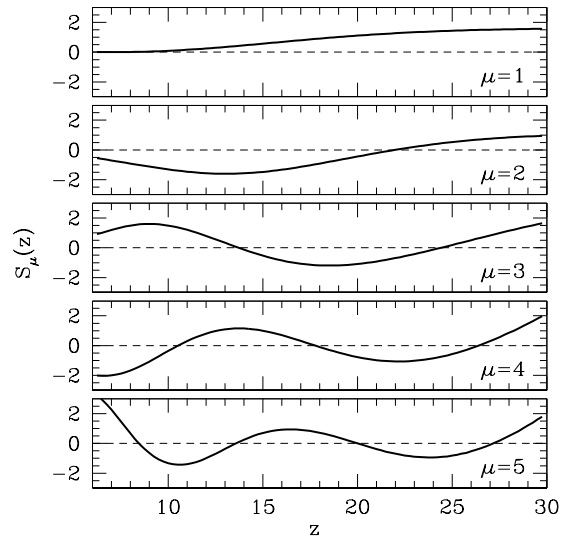, width=3.0in}}
\caption{The five lowest-variance principal components of $x_e(z)$ 
over redshifts $6<z<30$, with increasing variance from top to bottom.
%\vskip 0.25cm
}
\label{fig:pc}
\end{figure}
% ****************************************

For complete representation of the effects of 
reionization between $\zmin$ and $\zmax$ on the low-$\ell$ $E$-mode 
spectrum to better accuracy than 
cosmic variance, no more than the first five principal components are 
needed (assuming $\zmax\lesssim 30$)~\cite{HuHol03,MorHu07b}. 
Due to projection effects~\cite{HuWhi97c}, 
the accuracy to which the lowest-variance principal components reconstruct 
$\cltens$ at low $\ell$ is even better than for the scalar $\clee$.
In the MCMC analysis presented in the following section, we always use 
the five lowest-variance principal components of $x_e(z)$ with $\zmax=30$, which 
we show in Fig.~\ref{fig:pc} (see~\cite{MorHu07b} for 
the effects of using a different number of components to analyze $E$-mode data). 
The amplitudes of these components then serve to 
parametrize general reionization histories in the analysis of 
CMB polarization data.

% =====================================================
\section{Markov Chain Monte Carlo Constraints}
\label{sec:mcmc}

We use Markov Chain Monte Carlo (MCMC) analysis to explore the joint 
effects of the reionization history and inflationary 
parameters on CMB polarization power spectra~\cite[see e.g.][]{Chretal01,KosMilJim02,Dunetal04}. 
Chains of Monte Carlo samples are generated using the publicly available 
code CosmoMC~\cite{LewBri02} \footnote{{\tt http://cosmologist.info/cosmomc/}},
which includes the code CAMB~\citep{Lewetal00} for computing 
theoretical angular power spectra at each point in the  
parameter space. We have modified both codes to allow 
specification of an arbitrary reionization history calculated from a set 
of principal component amplitudes using Eq.~(\ref{eq:mmutoxe}), as described in~\cite{MorHu07b}. 

For chains in which $x_e(z)$ is parametrized by its principal components, 
the parameters that we vary include the amplitudes of the first five components, 
the tensor-to-scalar ratio at $k_{\rm pivot}=0.05$~Mpc$^{-1}$, and the tilt of the tensor spectrum:
$\{m_1,m_2,m_3,m_4,m_5,r_{0.05},n_t\}$. When using the simple instantaneous 
form of $x_e(z)$, the chain parameters are instead $\{\tau,r_{0.05},n_t\}$ with 
optical depth taking the place of principal components. 
We compute $r$ at $k_{\rm pivot}=0.01$~Mpc$^{-1}$ as a derived parameter using 
Eq.~(\ref{eq:rscale}) after generating each chain.

When analyzing the parameter chains, we impose priors on the 
principal component amplitudes corresponding to physical values of the 
ionized fraction, $0 \leq x_e \leq 1$~\cite{MorHu07b}. 
These priors are conservative in the sense that all excluded models are unphysical, 
but the models we retain are not necessarily physical.

We use only polarization for parameter constraints, 
and assume that the values of the 
standard \lcdm\ parameters (besides $\tau$) are fixed by measurements of the 
CMB temperature anisotropies. 
To a good approximation the effect of $x_e(z)$ on the large-scale 
polarization is independent of the other parameters~\citep{MorHu07b}. 

Specifically, we take  
$\Omega_bh^2=0.0222$, $\Omega_ch^2=0.106$, $100\theta=1.04$ (corresponding 
to $h=0.73$), $A_s e^{-2\tau}=1.7 \times 10^{-9}$, and $n_s=0.96$, 
consistent with the \wmap\ 3-year temperature power spectrum.
When computing the optical depth to reionization we 
take the  helium fraction to be $Y_p=0.24$ and assume that helium is neutral. 
With these parameter values, the reionization optical depth out to $\zmin = 6$ 
is fixed at $\tau(\zmin)\approx 0.04$. 
The remaining contribution to the total optical depth from $\zmin < z < \zmax$ 
is determined by the 
values of $\{m_{\mu}\}$ for each sample in the chains.
The default bin width for the fiducial models and principal components 
is $\dz=0.25$, which is small enough that 
numerical effects related to binning should be negligible.

For each scenario that we study, we run four separate chains 
until the Gelman and Rubin convergence statistic $R$, 
corresponding to the ratio of the variance of parameters between chains to 
the variance within each chain, satisfies 
$R-1<0.01$~\citep{GelRub92,BroGel98}. The convergence diagnostic  
of~\cite{RafLew92} is used to determine how much each chain must be thinned 
to obtain independent samples.

% ****************************************
\begin{table*}
\caption{Constraints on $\tau$, $r$, and $n_t$ for simulated data based 
on the double reionization history with optical depth $\tau^{\rm fid}=0.09$.
}
\begin{center}
\begin{tabular}{lllllrlr}
\hline
\hline
 & & & & Use PCs &  &  &  \\
\multicolumn{1}{c}{$r_{0.05}^{\rm fid}$} & \multicolumn{1}{c}{$r^{\rm fid}$} & \multicolumn{1}{c}{$n_t^{\rm fid}$} & Noise & of $x_e(z)$? & \multicolumn{1}{c}{$\tau$} & \multicolumn{1}{c}{$r$} & \multicolumn{1}{c}{$n_t$} \\
\hline
0.3 & 0.299 & $-0.0375$ & CV & Yes & $0.091\pm0.005$ & $0.299\pm0.004$ & $-0.037\pm0.018$ \\
0.3 & 0.299 & $-0.0375$ & CV & No & $0.118\pm0.003$ & $0.298\pm0.005$ & $-0.021\pm0.018$ \\
0.3 & 0.299 & $-0.0375$ & Planck & Yes & $0.094\pm0.009$ & $0.321\pm0.095$ & $0.041\pm0.196$ \\
0.3 & 0.299 & $-0.0375$ & Planck & No & $0.113\pm0.007$ & $0.390\pm0.096$ & $0.398\pm0.164$ \\
0.03 & 0.0283 & $-0.00375$ & CV & Yes & $0.091\pm0.005$ & $0.0283\pm0.0006$ & $0.001\pm0.051$ \\
0.03 & 0.0283 & $-0.00375$ & CV & No & $0.121\pm0.003$ & $0.0288\pm0.0006$ & $0.162\pm0.049$ \\
0.03 & 0.0283 & $-0.00375$ & CMBPol & Yes & $0.092\pm0.007$ & $0.032\pm0.009$ & $0.069\pm0.184$ \\
0.03 & 0.0283 & $-0.00375$ & CMBPol & No & $0.122\pm0.005$ & $0.038\pm0.010$ & $0.491\pm0.165$ \\
\hline
\hline
\end{tabular}
\end{center}
\label{tab:rnttau}
\end{table*}
% ****************************************

% =====================================================
\subsection{Simulated data}
\label{sec:data}

We use CAMB to generate model $E$- and $B$-mode polarization 
power spectra. We take the data to be the exact values of a given model
$\hat\clee =\clee$ and $\hat\clbb =\clbb$, 
neglecting cosmic variance and noise, so we expect constraints to be 
centered on the fiducial parameter values rather than displaced by 
$\sim 1~\sigma$. These constraints can be thought of as the average over 
many possible realizations of the data. (See~\cite{MorHu07b} for a discussion of the  
effects of cosmic variance when using realizations drawn from $C_{\ell}$ instead of 
taking $C_{\ell}$ as the data.)

For the $j$th sample in a chain, the likelihood is
\begin{eqnarray}
-\ln L_{(j)}&=&\sum_{\ell=2}^{\lmax}\left(\ell+\frac{1}{2}\right)
f_{\rm sky}^2 \\
&& \times
\left(\frac{\hat{{\bf C}}_{\ell}^{EE}}{{\bf C}_{\ell(j)}^{EE}}
+\frac{\hat{{\bf C}}_{\ell}^{BB}}{{\bf C}_{\ell(j)}^{BB}}
+\ln\frac{{\bf C}_{\ell(j)}^{EE}{\bf C}_{\ell(j)}^{BB}}
{\hat{{\bf C}}_{\ell}^{EE}\hat{{\bf C}}_{\ell}^{BB}}-2\right), \nonumber
\label{eq:like}
\end{eqnarray}
where 
$\hat{{\bf C}}_{\ell}=\hat{C}_{\ell}+N_{\ell}$ is the sum of the 
simulated data and noise spectra, and 
${\bf C}_{\ell(j)}=C_{\ell(j)}+N_{\ell}$ is the 
theoretical spectrum calculated with the 
parameter values at step $j$ in the chain plus the noise spectrum. 
For consistency with
CosmoMC, the likelihood contains an extra factor of $f_{\rm sky}$, 
the fraction of sky observed \footnote{The extra factor
of $f_{\rm sky}$ is meant to approximately model the loss of information due to mode coupling 
with incomplete sky coverage, but the correct approach depends on the nature of the information extracted.}.

We set $N_{\ell}=0$ to simulate measurements limited only by cosmic variance.
For more realistic scenarios based on \planck\ and \cmbpol, we model noise as
\begin{equation}
N_{\ell} = \left(\frac{w_p^{-1/2}}{\mu{\rm K\mbox{-}rad}}\right)^2 
\exp\left[\frac{\ell(\ell+1)(\theta_{\rm FWHM}/{\rm rad})^2}{8\ln 2}\right],
\end{equation}
where $w_p^{-1/2}$ is the polarization noise level and 
$\theta_{\rm FWHM}$ is the beam width.

We typically compute the likelihood up to $\lmax=1000$. The $E$-mode spectrum 
and $B$-mode tensor spectrum are fixed at $\ell \gtrsim 100$ by setting 
$A_s e^{-2\tau}$ constant in the Monte Carlo chains. The impact on 
$r$ and $n_t$ constraints of the $B$-mode spectrum at multipoles 
greater than a few hundred is negligible in the presence of the 
lensing spectrum and experimental noise for all allowed values of $r$.

The fiducial reionization history is the extended, double reionization model with 
polarization spectra plotted in Fig.~\ref{fig:cl}, for which $\tau=0.090$. 
This function is chosen because it is not well described by the instantaneous 
reionization model, so biases due to assuming instantaneous $x_e(z)$ 
should be readily apparent in the parameter constraints.

We consider two fiducial values of the tensor-to-scalar 
ratio, $r_{0.05}=0.3$ and $r_{0.05}=0.03$. Using the consistency relation [Eq.~(\ref{eq:consrel})] 
to set $n_t$ and taking the fiducial scalar tilt $n_s=0.96$, these tensor-to-scalar 
ratios correspond to $r=0.299$ and $r=0.0283$ at $k_{\rm pivot}=0.01$~Mpc$^{-1}$, respectively.

Assuming a power-law primordial spectrum, the larger fiducial tensor-to-scalar 
ratio, $r_{0.05}=0.3$, is approximately the 68\% 
upper limit on $r$ from 3-year \wmap\ data, although 
when large-scale structure data are included it is ruled out at 
about 95\% CL~\cite{Speetal07}.
A tensor spectrum with this value of $r$ should be 
detectable by the \planck\ satellite~\cite{Planck}. 
The smaller value of $r$ may be out of 
the reach of \planck\ but accessible to a next-generation CMB satellite, 
such as the proposed \cmbpol\ \cite{CMBTaskForce}. 
In addition to CV-limited 
measurements, we consider a noise spectrum based on 
\planck\ with sensitivity $w_p^{-1/2}=81~\mu{\rm K}'$ and beam size 
$\theta_{\rm FWHM}=7.1'$ for the $r_{0.05}=0.3$ simulated data~\cite{Planck,Albetal06}, and 
low-resolution \cmbpol-like noise with $w_p^{-1/2}=20~\mu{\rm K}'$ and
$\theta_{\rm FWHM}=60'$ for $r_{0.05}=0.03$~\cite{CMBTaskForce} 
(dotted curves in Fig.~\ref{fig:cl}).
We assume $f_{\rm sky}=0.8$ for both \planck\ and \cmbpol\ and take $f_{\rm sky}=1$ 
for the more idealized, CV-limited data.
In all cases we neglect foregrounds, assuming that they can be
adequately subtracted using polarization data from frequency channels 
not used for cosmological parameter estimation.

The $B$-mode lensing spectrum has a significant impact on constraints on $r$ and 
$n_t$, but including lensing significantly slows down the computation of 
the angular power spectra in CAMB. However, $\cllens$ is nearly independent of 
the parameters that we vary in the Monte Carlo chains. Rather than computing 
the effects of lensing directly, then, we treat the lensing 
spectrum as a fixed contribution to the noise power spectrum. 
Tests comparing this approximation to an analysis including the
full effects of lensing show that the constraints obtained for $r$ and $n_t$ are 
the same. Using $\tau$-dependent lensing spectra computed for each 
Monte Carlo sample appears to improve the constraint on $\tau$, but this is 
an artifact due to fixing $A_s e^{-2\tau}$ and other parameters that also
affect $\cllens$, mainly  $\Omega_m h^2$~\cite{SmiHuKap06}.

% ****************************************
\begin{figure}
\centerline{\psfig{file=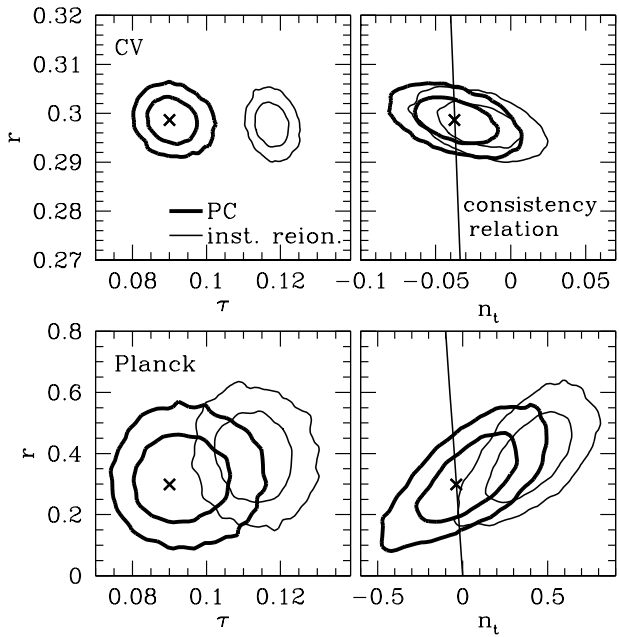, width=3.5in}}
\caption{2D marginalized 68 and 95\% contours for the optical depth ($\tau$),
tensor-to-scalar ratio ($r$), and tensor spectral index ($n_t$). 
For the simulated polarization spectra, we take $x_e(z)$ to be a double  
reionization history with $\tau=0.090$ (see Fig.~\ref{fig:cl}).
The fiducial tensor-to-scalar ratio is set to $r_{0.05}=0.3$ ($r=0.299$ at $k=0.01$~Mpc$^{-1}$), 
and the fiducial tensor tilt is assumed 
to obey the consistency relation, $n_t=-r/8$ (\emph{right panels, line}). 
Crosses indicate these fiducial parameter values. 
For the thick contours, the chains include the five 
lowest-variance principal components of $x_e(z)$.  Thin contours 
show the constraints when $x_e(z)$ is instead assumed to be 
instantaneous and parametrized only by $\tau$.
The lensing spectrum is treated as a contribution to the noise.  
\emph{Top panels}: CV-limited data; \emph{bottom}: including \planck-like noise. 
}
\label{fig:rnttau03}
\end{figure}
% ****************************************

% ****************************************
\begin{figure}
\centerline{\psfig{file=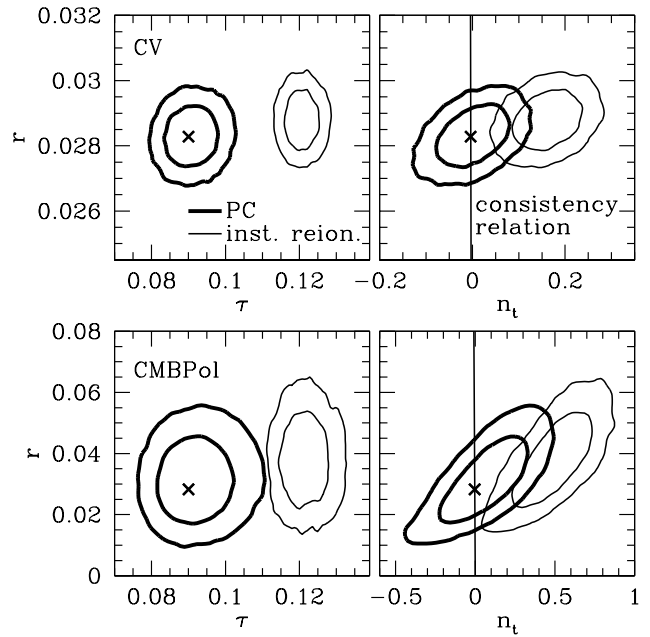, width=3.5in}}
\caption{Same as Fig.~\ref{fig:rnttau03}, but with fiducial 
tensor-to-scalar ratio $r_{0.05}=0.03$ ($r=0.0283$ at $k=0.01$~Mpc$^{-1}$). 
\emph{Top panels}: CV-limited data; 
\emph{bottom}: including \cmbpol-like noise.
}
\label{fig:rnttau003}
\end{figure}
% ****************************************

% =====================================================
\subsection{Results}
\label{sec:results}

Table~\ref{tab:rnttau} lists the 1D marginalized constraints on 
$\tau$, $r$, and $n_t$ from the MCMC 
analysis for each case study. 
The constraints on $\tau$, $r$, and $n_t$ are plotted in 
Figs.~\ref{fig:rnttau03} and~\ref{fig:rnttau003} as 2D contours 
after marginalizing over all other parameters. The two sets of contours in each panel
use the same simulated data, but different parametrizations of $x_e(z)$. 
For the thick contours, the five lowest-variance principal components of $x_e(z)$ 
are included in the Monte Carlo chains, while the thin contours come from 
chains that treat $x_e(z)$ as instantaneous reionization with only one 
parameter, $\tau$. For the principal component chains, $\tau$ is 
derived from the principal component amplitudes, $\{m_{\mu}\}$.

Since the constraint on optical depth comes primarily from the 
reionization peak
of the $E$-mode spectrum, estimates of $\tau$ in all cases in 
Tab.~\ref{tab:rnttau} are affected similarly by 
the parametrization of the reionization history. As noted in~\cite{Kapetal03,LewWelBat06}, using the 
one-parameter instantaneous $x_e(z)$ when that model is not sufficient to describe 
the true reionization history (in this case, the double reionization history 
represented by the thick curves in Fig.~\ref{fig:cl}) can lead to a significant 
bias in the constraint on $\tau$.  Using a set of principal components to 
parametrize $x_e(z)$ removes the bias in the optical depth~\cite{MorHu07b}.

The impact of the reionization history assumptions for inflationary parameters
depends more strongly on the values of $r$ and the noise. 
We consider first the most optimistic scenario for measurement of the 
$B$-mode tensor spectrum: a CV-limited measurement of 
$\cltens$ with approximately the largest amplitude currently allowed, 
$r_{0.05}=0.3$. With this tensor-to-scalar ratio, the tensor spectrum 
dominates over the $B$-mode lensing spectrum for $\ell \lesssim 150$ 
(see Fig.~\ref{fig:cl}), so the recombination peak of $\cltens$ is 
free from contamination over roughly a decade in $\ell$.  Moreover, the lensing
spectrum may be statistically subtracted well beyond the multipole where
it contributes equal power.

Since the amplitude of the tensor spectrum depends on $\tau$, $r$, and $n_t$, 
we expect some degeneracy between these parameters, as described in \S~\ref{sec:param}. 
Because of this degeneracy, a bias in $\tau$ can generate biases in the inflationary 
parameters as well, mainly $n_t$ (since $r$ at $k=0.01$~Mpc$^{-1}$ is tightly constrained 
by measuring $\cltens$ at $\ell\approx 100$).  However, $\tau$ only 
affects the low-$\ell$ reionization peak (with our assumption that 
$A_s e^{-2\tau}$ is held fixed), so these biases will only 
be significant if the reionization peak contributes significantly to 
constraints on the tensor spectrum. In this optimistic scenario with $r_{0.05}=0.3$ and 
cosmic variance-limited measurements, enough of the tensor recombination peak 
is observable so that additionally measuring
the reionization peak has a negligible effect on 
constraints on $r$ and $n_t$; the values of these parameters are 
determined almost entirely by $\clbb$ at $20 \lesssim \ell \lesssim 500$. 
As a result, the bias in the inflationary parameters 
due to incorrectly assuming instantaneous reionization is small; as 
the contours in the top right panel of Fig.~\ref{fig:rnttau03} show, the true parameter values 
remain within the 68\% CL even with this bias.

Now consider the same tensor-to-scalar ratio but more realistic assumptions 
about the experimental noise. Using our assumed noise spectrum for \planck\ 
as described in \S~\ref{sec:data}, we find constraints on $\tau$, $r$, and $n_t$ 
as shown in the bottom panels of Fig.~\ref{fig:rnttau03}.
The most obvious difference from the CV-limited case is that 
the parameter uncertainties are larger, especially for the inflationary 
parameters. There is also a difference in the instantaneous reionization biases: 
while the optical depth bias is similar to the CV-limited case, 
$n_t$ is biased much more with \planck-like noise than with a CV-limited measurement. 
The additional bias is due to the greater dependence on the reionization peak of $\cltens$ 
for $n_t$ constraints since the noise associated with \planck\ 
makes measurement of the recombination peak much more difficult.  

\begin{figure}
\centerline{\psfig{file=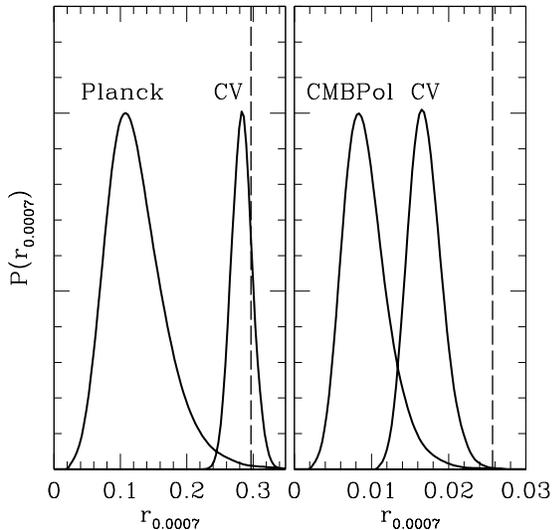, width=3.0in}}
\caption{1D marginalized distributions of $r_{0.0007}$, the tensor-to-scalar ratio on scales
near the reionization peak, for each 
of the four MCMC scenarios in Figs.~\ref{fig:rnttau03} and~\ref{fig:rnttau003}.
The fiducial values of $r_{0.0007}$ are indicated by vertical dashed lines, and the normalization
is arbitrary.
}
\label{fig:rbias}
\end{figure}

Note that while $r$ at $k_{\rm pivot}=0.01$~Mpc$^{-1}$ is still determined 
fairly accurately in this scenario, 
if the tensor-to-scalar ratio were quoted elsewhere, e.g. as $r_{0.002}$ or $r_{0.05}$, 
a significant bias would appear in that parameter as well.
For example, Fig.~\ref{fig:rbias} shows that $r_{0.0007}$, normalized at the 
approximate scale of the reionization peak, lies significantly 
below the fiducial value in most of our test cases.
The choice of scale also affects the uncertainty in the tensor-to-scalar ratio; 
while the error we quote for $n_t$ agrees with similar studies, e.g.~\cite{VerPeiJim06}, 
the error on $r$ differs in general since degeneracy with $n_t$ due to a 
different choice of $k_{\rm pivot}$ can greatly increase $\sigma_r$.

% ****************************************
% ****************************************

With the larger bias in $n_t$ in the presence of 
\planck-like noise, the true parameter values are excluded at the 
95\% CL. Moreover, the consistency relation for single-field slow-roll 
inflation (plotted as a line in the $r-n_t$ plots) 
is excluded at about the same confidence level. An incorrect assumption about 
the reionization history could therefore lead to wrongly rejecting the simplest
class of inflationary models. 

As the thick contours in Fig.~\ref{fig:rnttau03} show, 
when general reionization histories are considered by parametrizing 
$x_e(z)$ by its principal components, the biases in $\tau$, $r$, and $n_t$ are 
all removed. The constraints on $r$ and $n_t$ still may not be very strong for 
an experiment like \planck, but at least when using the principal components 
of $x_e(z)$ one would not be led to exclude the true underlying model 
of inflation.  Furthermore, the precision of the constraints is not substantially
degraded by the principal component parametrization.

Note that the exact magnitude and direction of the parameter biases depend 
on what the true reionization history actually is. The goal here is not to 
make specific predictions about these biases, 
but rather to give an idea of how large they
could be and to show that they can be avoided by allowing for 
general models of reionization.

How does uncertainty about reionization influence constraints on inflationary 
parameters if the amplitude of the tensor spectrum is smaller? In general, 
we expect the importance of the reionization peak of $\cltens$ relative 
to the recombination peak to increase as the tensor spectrum drops further 
below the dominant source of ``noise'' on large scales, whether that be the 
noise spectrum of an experiment or the $B$-mode lensing signal. 
To check this expectation, we use simulated data with a fiducial tensor-to-scalar 
ratio of $r_{0.05}=0.03$. The resulting constraints on $\tau$, $r$, and $n_t$, 
plotted as for the $r_{0.05}=0.3$ results, are shown in Fig.~\ref{fig:rnttau003} 
for both cosmic variance-limited measurements 
(top panels) and the \cmbpol-like noise spectrum described in 
\S~\ref{sec:data} (bottom panels).

With a CV-limited experiment and $r_{0.05}=0.03$, 
due to the greater influence of lensing relative to the tensor spectrum, 
the reionization peak ($\ell<20$) and recombination peak at $\ell\lesssim 300$ 
contribute roughly equally to the constraint on $n_t$. 
Because of this, assuming the wrong reionization history biases $n_t$ more 
than for $r_{0.05}=0.3$, and in this case 
the consistency relation would be excluded at more than the $95\%$ CL.

If we add the \cmbpol\ noise spectrum ($w_p^{-1/2}=20~\mu{\rm K}'$, 
$\theta_{\rm FWHM}=60'$), both biases and errors on 
parameters are larger. The constraints in this case are quite similar 
to the constraints in the case of {\it Planck} with $r_{0.05}=0.3$ (except with all 
$r$ values reduced by a factor of 10), which makes sense 
since the low-$\ell$ amplitude of \cmbpol\ noise relative to $\cltens$ with $r_{0.05}=0.03$ is
similar to that of \planck\ noise to $\cltens$ with $r_{0.05}=0.3$ (see Fig.~\ref{fig:cl}).
Note that our assumptions about the \cmbpol\ noise spectrum are 
on the conservative end of the range usually considered for such an 
experiment; better sensitivity ($\sim 1~\mu{\rm K}'$) and/or resolution 
($\sim 1-10$~arcmin) would enable measurement of the tensor $B$-modes to 
smaller scales so that constraints on inflationary parameters would 
be closer to the CV-limited scenario.

For even lower values of $r$ than those considered here, the impact 
of reionization on constraints on inflationary parameters is likely to 
be the same or greater. Even for an idealized, cosmic variance-limited 
experiment, the $B$-mode signal due to lensing becomes a significant 
contaminant of $\cltens$ at $r<0.03$.  Due to the shape of the spectra, if 
any part of $\cltens$ is detectable above the lensing $B$-modes and noise 
for low tensor-to-scalar ratios it will be the reionization peak. 
Any bias in $\tau$ due to incorrect modeling of the reionization history 
will then cause the inferred value of $n_t$ to be biased. 
Of course, at very low $r$ the uncertainties are so large that
the parameters are not usefully constrained at all.
In this case, it may be possible to measure $n_t$ using 
direct observations by future gravitational-wave experiments to better 
accuracy than what is possible with CMB polarization~\cite{SmiPeiCoo06,ChoEfs06}.

Finally, note that although the choice of normalization scale $k_{\rm pivot}=0.01$~Mpc$^{-1}$ 
is intended to decorrelate $r$ from the other parameters, there is some remaining 
degeneracy with $n_t$. The direction of this degeneracy depends on whether or not 
$\cltens$ can be measured on scales smaller than $k_{\rm pivot}$. For CV-limited 
measurement of a $B$-mode spectrum with $r=0.3$, these smaller scales are 
observable. A larger tensor tilt that increases the power at $\ell \gtrsim 100$ 
can be compensated for by lowering the value of $r$, so $r$ and $n_t$ are anticorrelated 
in this case (top right panel of Fig.~\ref{fig:rnttau03}). 
If the tensor spectrum on such small scales is hidden by 
lensing $B$-modes or noise, then the low-$\ell$ side of the recombination peak becomes more 
important. Increasing tensor tilt lowers the power at $20 \lesssim \ell \lesssim 100$, 
so $r$ is correlated with $n_t$ so as to match the spectrum of the data on these scales 
(lower right panel of Fig.~\ref{fig:rnttau03} and right panels of Fig.~\ref{fig:rnttau003}).

% =====================================================
\section{Discussion}
\label{sec:discuss}

The value of the optical depth to reionization estimated from the 
CMB $E$-mode polarization spectrum on large scales can be biased by 
adopting a model that has insufficient freedom to 
describe the true reionization history.  Likewise, the use of 
simple reionization models can bias inflationary parameters such 
as the tensor-to-scalar ratio and tensor tilt that depend on the 
large-scale amplitude of the $B$-mode spectrum of primordial 
gravitational waves. In each case, the problem can be solved by 
using a more general parametrization of the reionization history. 
We have shown that using a small but complete set of the 
principal components of the reionization history effectively 
yields unbiased constraints on both reionization and inflationary parameters.

Measurements of $r$ and $n_t$ are only affected by 
the assumed form of the reionization history 
if the reionization peak of the tensor $B$-mode spectrum 
at the very largest scales is needed to precisely constrain the
parameters.  If, instead, good constraints can 
be obtained using only the $B$-mode recombination peak at intermediate scales, 
then assumptions about reionization do not affect tests of the 
consistency relation between $r$ and $n_t$.  They would
instead appear as false evidence for running of the tensor tilt in violation of slow-roll
expectations.
Measurement of the recombination peak however is inhibited by experimental noise and contamination from $E$-mode power converted to $B$-mode power 
by gravitational lensing, 
both of which become more important at smaller scales.

To study the potential impact of reionization on 
parameter constraints from $B$-mode polarization,
we have employed a Markov Chain Monte Carlo analysis of simulated 
CMB polarization power spectra and compared results for two 
descriptions of reionization: a simple, one-parameter, instantaneous 
reionization model, and a parametrization using 
principal components of the reionization history 
with respect to the $E$-mode polarization power spectrum. 

By varying the properties of the simulated polarization power spectra, including 
the fiducial tensor-to-scalar ratio and the noise spectrum, we 
have determined over what range of scales CMB polarization data is most important 
for constraining inflationary parameters in various scenarios. 
In particular, the question of whether the large-scale reionization peak 
of $\cltens$ or the smaller-scale 
recombination peak is more important determines the severity of bias in 
inflationary parameters when reionization is modeled incorrectly. 

If the tensor-to-scalar ratio is near the current upper limit of $r\sim 0.3$ 
and measurements of $B$-mode polarization are limited only by cosmic 
variance, then the spectrum on scales $20 \lesssim \ell \lesssim 500$ 
dominates constraints on $r$ and $n_t$ and incorrect assumptions about 
reionization do not strongly bias the results. If the true tensor-to-scalar 
ratio is more than a factor of a few smaller than this upper bound, however, 
then lensing $B$-modes limit the information that can be extracted from 
the recombination peak of the tensor spectrum alone. Furthermore, all-sky
experiments in the foreseeable future are likely to have noise that 
exceeds the lensing signal, making tests of inflation even more 
reliant on the reionization peak of the tensor $B$-modes on large scales.
In all of these cases, a general parametrization of reionization 
such as that provided by principal components allows the use of the 
$B$-mode reionization peak for inflationary parameter constraints without 
significantly worsening the errors on those parameters.

\vfill
\acknowledgments 
We thank S. DeDeo, C. Dvorkin, H. Peiris, A. Upadhye, and S. Wang
for useful conversations.
This work was supported by the KICP through the grant NSF PHY-0114422 and
the David and
Lucile Packard Foundation. 
MJM was additionally supported 
by a 
National Science Foundation Graduate Research Fellowship. 
WH was additionally supported by  the DOE through 
contract DE-FG02-90ER-40560.

\bibliographystyle{apsrev}

\begin{thebibliography}{54}
\expandafter\ifx\csname natexlab\endcsname\relax\def\natexlab#1{#1}\fi
\expandafter\ifx\csname bibnamefont\endcsname\relax
  \def\bibnamefont#1{#1}\fi
\expandafter\ifx\csname bibfnamefont\endcsname\relax
  \def\bibfnamefont#1{#1}\fi
\expandafter\ifx\csname citenamefont\endcsname\relax
  \def\citenamefont#1{#1}\fi
\expandafter\ifx\csname url\endcsname\relax
  \def\url#1{\texttt{#1}}\fi
\expandafter\ifx\csname urlprefix\endcsname\relax\def\urlprefix{URL }\fi
\providecommand{\bibinfo}[2]{#2}
\providecommand{\eprint}[2][]{\url{#2}}

\bibitem[{\citenamefont{{Guth}}(1981)}]{Gut81}
\bibinfo{author}{\bibfnamefont{A.~H.} \bibnamefont{{Guth}}},
  \bibinfo{journal}{\prd} \textbf{\bibinfo{volume}{23}}, \bibinfo{pages}{347}
  (\bibinfo{year}{1981}).

\bibitem[{\citenamefont{{Albrecht} and {Steinhardt}}(1982)}]{AlbSte82}
\bibinfo{author}{\bibfnamefont{A.}~\bibnamefont{{Albrecht}}} \bibnamefont{and}
  \bibinfo{author}{\bibfnamefont{P.~J.} \bibnamefont{{Steinhardt}}},
  \bibinfo{journal}{Physical Review Letters} \textbf{\bibinfo{volume}{48}},
  \bibinfo{pages}{1220} (\bibinfo{year}{1982}).

\bibitem[{\citenamefont{{Linde}}(1982)}]{Lin82}
\bibinfo{author}{\bibfnamefont{A.~D.} \bibnamefont{{Linde}}},
  \bibinfo{journal}{Physics Letters B} \textbf{\bibinfo{volume}{108}},
  \bibinfo{pages}{389} (\bibinfo{year}{1982}).

\bibitem[{\citenamefont{{Sato}}(1981)}]{Sat81}
\bibinfo{author}{\bibfnamefont{K.}~\bibnamefont{{Sato}}},
  \bibinfo{journal}{\mnras} \textbf{\bibinfo{volume}{195}},
  \bibinfo{pages}{467} (\bibinfo{year}{1981}).

\bibitem[{\citenamefont{{Hu} and {White}}(1996)}]{HuWhi96c}
\bibinfo{author}{\bibfnamefont{W.}~\bibnamefont{{Hu}}} \bibnamefont{and}
  \bibinfo{author}{\bibfnamefont{M.}~\bibnamefont{{White}}},
  \bibinfo{journal}{\prl} \textbf{\bibinfo{volume}{77}}, \bibinfo{pages}{1687}
  (\bibinfo{year}{1996}), \eprint{arXiv:astro-ph/9602020}.

\bibitem[{\citenamefont{{Spergel} and {Zaldarriaga}}(1997)}]{SpeZal97}
\bibinfo{author}{\bibfnamefont{D.~N.} \bibnamefont{{Spergel}}}
  \bibnamefont{and}
  \bibinfo{author}{\bibfnamefont{M.}~\bibnamefont{{Zaldarriaga}}},
  \bibinfo{journal}{Physical Review Letters} \textbf{\bibinfo{volume}{79}},
  \bibinfo{pages}{2180} (\bibinfo{year}{1997}),
  \eprint{arXiv:astro-ph/9705182}.

\bibitem[{\citenamefont{{Hu} et~al.}(1997)\citenamefont{{Hu}, {Spergel}, and
  {White}}}]{HuSpeWhi97}
\bibinfo{author}{\bibfnamefont{W.}~\bibnamefont{{Hu}}},
  \bibinfo{author}{\bibfnamefont{D.~N.} \bibnamefont{{Spergel}}},
  \bibnamefont{and} \bibinfo{author}{\bibfnamefont{M.}~\bibnamefont{{White}}},
  \bibinfo{journal}{\prd} \textbf{\bibinfo{volume}{51}}, \bibinfo{pages}{3288}
  (\bibinfo{year}{1997}), \eprint{arXiv:astro-ph/9605193}.

\bibitem[{\citenamefont{{Peiris} et~al.}(2003)\citenamefont{{Peiris},
  {Komatsu}, {Verde}, {Spergel}, {Bennett}, {Halpern}, {Hinshaw}, {Jarosik},
  {Kogut}, {Limon} et~al.}}]{Peietal03}
\bibinfo{author}{\bibfnamefont{H.~V.} \bibnamefont{{Peiris}}},
  \bibinfo{author}{\bibfnamefont{E.}~\bibnamefont{{Komatsu}}},
  \bibinfo{author}{\bibfnamefont{L.}~\bibnamefont{{Verde}}},
  \bibinfo{author}{\bibfnamefont{D.~N.} \bibnamefont{{Spergel}}},
  \bibinfo{author}{\bibfnamefont{C.~L.} \bibnamefont{{Bennett}}},
  \bibinfo{author}{\bibfnamefont{M.}~\bibnamefont{{Halpern}}},
  \bibinfo{author}{\bibfnamefont{G.}~\bibnamefont{{Hinshaw}}},
  \bibinfo{author}{\bibfnamefont{N.}~\bibnamefont{{Jarosik}}},
  \bibinfo{author}{\bibfnamefont{A.}~\bibnamefont{{Kogut}}},
  \bibinfo{author}{\bibfnamefont{M.}~\bibnamefont{{Limon}}},
  \bibnamefont{et~al.}, \bibinfo{journal}{\apjs}
  \textbf{\bibinfo{volume}{148}}, \bibinfo{pages}{213} (\bibinfo{year}{2003}),
  \eprint{arXiv:astro-ph/0302225}.

\bibitem[{\citenamefont{{Spergel} et~al.}(2007)\citenamefont{{Spergel}, {Bean},
  {Dor{\'e}}, {Nolta}, {Bennett}, {Dunkley}, {Hinshaw}, {Jarosik}, {Komatsu},
  {Page} et~al.}}]{Speetal07}
\bibinfo{author}{\bibfnamefont{D.~N.} \bibnamefont{{Spergel}}},
  \bibinfo{author}{\bibfnamefont{R.}~\bibnamefont{{Bean}}},
  \bibinfo{author}{\bibfnamefont{O.}~\bibnamefont{{Dor{\'e}}}},
  \bibinfo{author}{\bibfnamefont{M.~R.} \bibnamefont{{Nolta}}},
  \bibinfo{author}{\bibfnamefont{C.~L.} \bibnamefont{{Bennett}}},
  \bibinfo{author}{\bibfnamefont{J.}~\bibnamefont{{Dunkley}}},
  \bibinfo{author}{\bibfnamefont{G.}~\bibnamefont{{Hinshaw}}},
  \bibinfo{author}{\bibfnamefont{N.}~\bibnamefont{{Jarosik}}},
  \bibinfo{author}{\bibfnamefont{E.}~\bibnamefont{{Komatsu}}},
  \bibinfo{author}{\bibfnamefont{L.}~\bibnamefont{{Page}}},
  \bibnamefont{et~al.}, \bibinfo{journal}{\apjs}
  \textbf{\bibinfo{volume}{170}}, \bibinfo{pages}{377} (\bibinfo{year}{2007}),
  \eprint{arXiv:astro-ph/0603449}.

\bibitem[{\citenamefont{{Kamionkowski}
  et~al.}(1997)\citenamefont{{Kamionkowski}, {Kosowsky}, and
  {Stebbins}}}]{KamKosSte97}
\bibinfo{author}{\bibfnamefont{M.}~\bibnamefont{{Kamionkowski}}},
  \bibinfo{author}{\bibfnamefont{A.}~\bibnamefont{{Kosowsky}}},
  \bibnamefont{and}
  \bibinfo{author}{\bibfnamefont{A.}~\bibnamefont{{Stebbins}}},
  \bibinfo{journal}{Physical Review Letters} \textbf{\bibinfo{volume}{78}},
  \bibinfo{pages}{2058} (\bibinfo{year}{1997}),
  \eprint{arXiv:astro-ph/9609132}.

\bibitem[{\citenamefont{{Seljak} and {Zaldarriaga}}(1997)}]{SelZal97}
\bibinfo{author}{\bibfnamefont{U.}~\bibnamefont{{Seljak}}} \bibnamefont{and}
  \bibinfo{author}{\bibfnamefont{M.}~\bibnamefont{{Zaldarriaga}}},
  \bibinfo{journal}{Physical Review Letters} \textbf{\bibinfo{volume}{78}},
  \bibinfo{pages}{2054} (\bibinfo{year}{1997}),
  \eprint{arXiv:astro-ph/9609169}.

\bibitem[{\citenamefont{{The Planck Collaboration}}(2006)}]{Planck}
\bibinfo{author}{\bibnamefont{{The Planck Collaboration}}}
  (\bibinfo{year}{2006}), \eprint{arXiv:astro-ph/0604069}.

\bibitem[{\citenamefont{{Oxley} et~al.}(2004)\citenamefont{{Oxley}, {Ade},
  {Baccigalupi}, {deBernardis}, {Cho}, {Devlin}, {Hanany}, {Johnson}, {Jones},
  {Lee} et~al.}}]{Oxletal04}
\bibinfo{author}{\bibfnamefont{P.}~\bibnamefont{{Oxley}}},
  \bibinfo{author}{\bibfnamefont{P.~A.} \bibnamefont{{Ade}}},
  \bibinfo{author}{\bibfnamefont{C.}~\bibnamefont{{Baccigalupi}}},
  \bibinfo{author}{\bibfnamefont{P.}~\bibnamefont{{deBernardis}}},
  \bibinfo{author}{\bibfnamefont{H.-M.} \bibnamefont{{Cho}}},
  \bibinfo{author}{\bibfnamefont{M.~J.} \bibnamefont{{Devlin}}},
  \bibinfo{author}{\bibfnamefont{S.}~\bibnamefont{{Hanany}}},
  \bibinfo{author}{\bibfnamefont{B.~R.} \bibnamefont{{Johnson}}},
  \bibinfo{author}{\bibfnamefont{T.}~\bibnamefont{{Jones}}},
  \bibinfo{author}{\bibfnamefont{A.~T.} \bibnamefont{{Lee}}},
  \bibnamefont{et~al.}, in \emph{\bibinfo{booktitle}{Infrared Spaceborne Remote
  Sensing XII, Proceedings of the SPIE}}, edited by
  \bibinfo{editor}{\bibfnamefont{M.}~\bibnamefont{{Strojnik}}}
  (\bibinfo{year}{2004}), vol. \bibinfo{volume}{5543}, pp.
  \bibinfo{pages}{320--331}, \eprint{arXiv:astro-ph/0501111}.

\bibitem[{\citenamefont{{Yoon} et~al.}(2006)\citenamefont{{Yoon}, {Ade},
  {Barkats}, {Battle}, {Bierman}, {Bock}, {Brevik}, {Chiang}, {Crites},
  {Dowell} et~al.}}]{Yooetal06}
\bibinfo{author}{\bibfnamefont{K.~W.} \bibnamefont{{Yoon}}},
  \bibinfo{author}{\bibfnamefont{P.~A.~R.} \bibnamefont{{Ade}}},
  \bibinfo{author}{\bibfnamefont{D.}~\bibnamefont{{Barkats}}},
  \bibinfo{author}{\bibfnamefont{J.~O.} \bibnamefont{{Battle}}},
  \bibinfo{author}{\bibfnamefont{E.~M.} \bibnamefont{{Bierman}}},
  \bibinfo{author}{\bibfnamefont{J.~J.} \bibnamefont{{Bock}}},
  \bibinfo{author}{\bibfnamefont{J.~A.} \bibnamefont{{Brevik}}},
  \bibinfo{author}{\bibfnamefont{H.~C.} \bibnamefont{{Chiang}}},
  \bibinfo{author}{\bibfnamefont{A.}~\bibnamefont{{Crites}}},
  \bibinfo{author}{\bibfnamefont{C.~D.} \bibnamefont{{Dowell}}},
  \bibnamefont{et~al.}, in \emph{\bibinfo{booktitle}{Millimeter and
  Submillimeter Detectors and Instrumentation for Astronomy III, Proceedings of
  the SPIE}}, edited by
  \bibinfo{editor}{\bibfnamefont{J.}~\bibnamefont{{Zmuidzinas}}},
  \bibinfo{editor}{\bibfnamefont{W.~S.} \bibnamefont{{Holland}}},
  \bibinfo{editor}{\bibfnamefont{S.}~\bibnamefont{{Withington}}},
  \bibnamefont{and} \bibinfo{editor}{\bibfnamefont{W.~D.}
  \bibnamefont{{Duncan}}} (\bibinfo{year}{2006}), vol. \bibinfo{volume}{6275},
  \eprint{arXiv:astro-ph/0606278}.

\bibitem[{\citenamefont{Bock et~al.}(2006)}]{CMBTaskForce}
\bibinfo{author}{\bibfnamefont{J.}~\bibnamefont{Bock}} \bibnamefont{et~al.}
  (\bibinfo{year}{2006}), \eprint{arXiv:astro-ph/0604101}.

\bibitem[{\citenamefont{{Maffei} et~al.}(2005)\citenamefont{{Maffei}, {Ade},
  {Calderon}, {Challinor}, {de Bernardis}, {Dunlop}, {Gear},
  {Giraud-H{\'e}raud}, {Goldie}, {Grainge} et~al.}}]{Mafetal05}
\bibinfo{author}{\bibfnamefont{B.}~\bibnamefont{{Maffei}}},
  \bibinfo{author}{\bibfnamefont{P.~A.~R.} \bibnamefont{{Ade}}},
  \bibinfo{author}{\bibfnamefont{C.}~\bibnamefont{{Calderon}}},
  \bibinfo{author}{\bibfnamefont{A.~D.} \bibnamefont{{Challinor}}},
  \bibinfo{author}{\bibfnamefont{P.}~\bibnamefont{{de Bernardis}}},
  \bibinfo{author}{\bibfnamefont{L.}~\bibnamefont{{Dunlop}}},
  \bibinfo{author}{\bibfnamefont{W.~K.} \bibnamefont{{Gear}}},
  \bibinfo{author}{\bibfnamefont{Y.}~\bibnamefont{{Giraud-H{\'e}raud}}},
  \bibinfo{author}{\bibfnamefont{D.~J.} \bibnamefont{{Goldie}}},
  \bibinfo{author}{\bibfnamefont{K.~J.~B.} \bibnamefont{{Grainge}}},
  \bibnamefont{et~al.}, in \emph{\bibinfo{booktitle}{EAS Publications Series}}
  (\bibinfo{year}{2005}), pp. \bibinfo{pages}{251--256}.

\bibitem[{\citenamefont{{Lawrence} et~al.}(2004)\citenamefont{{Lawrence},
  {Gaier}, and {Seiffert}}}]{LawGaiSei04}
\bibinfo{author}{\bibfnamefont{C.~R.} \bibnamefont{{Lawrence}}},
  \bibinfo{author}{\bibfnamefont{T.~C.} \bibnamefont{{Gaier}}},
  \bibnamefont{and}
  \bibinfo{author}{\bibfnamefont{M.}~\bibnamefont{{Seiffert}}}, in
  \emph{\bibinfo{booktitle}{Millimeter and Submillimeter Detectors for
  Astronomy II, Proceedings of the SPIE}}, edited by
  \bibinfo{editor}{\bibfnamefont{C.~M.} \bibnamefont{{Bradford}}},
  \bibinfo{editor}{\bibfnamefont{P.~A.~R.} \bibnamefont{{Ade}}},
  \bibinfo{editor}{\bibfnamefont{J.~E.} \bibnamefont{{Aguirre}}},
  \bibinfo{editor}{\bibfnamefont{J.~J.} \bibnamefont{{Bock}}},
  \bibinfo{editor}{\bibfnamefont{M.}~\bibnamefont{{Dragovan}}},
  \bibinfo{editor}{\bibfnamefont{L.}~\bibnamefont{{Duband}}},
  \bibinfo{editor}{\bibfnamefont{L.}~\bibnamefont{{Earle}}},
  \bibinfo{editor}{\bibfnamefont{J.}~\bibnamefont{{Glenn}}},
  \bibinfo{editor}{\bibfnamefont{H.}~\bibnamefont{{Matsuhara}}},
  \bibinfo{editor}{\bibfnamefont{B.~J.} \bibnamefont{{Naylor}}},
  \bibnamefont{et~al.} (\bibinfo{year}{2004}), vol. \bibinfo{volume}{5498}, pp.
  \bibinfo{pages}{220--231}.

\bibitem[{\citenamefont{{MacTavish} et~al.}(2007)\citenamefont{{MacTavish},
  {Ade}, {Battistelli}, {Benton}, {Bihary}, {Bock}, {Bond}, {Brevik}, {Bryan},
  {Contaldi} et~al.}}]{Macetal07}
\bibinfo{author}{\bibfnamefont{C.~J.} \bibnamefont{{MacTavish}}},
  \bibinfo{author}{\bibfnamefont{P.~A.~R.} \bibnamefont{{Ade}}},
  \bibinfo{author}{\bibfnamefont{E.~S.} \bibnamefont{{Battistelli}}},
  \bibinfo{author}{\bibfnamefont{S.}~\bibnamefont{{Benton}}},
  \bibinfo{author}{\bibfnamefont{R.}~\bibnamefont{{Bihary}}},
  \bibinfo{author}{\bibfnamefont{J.~J.} \bibnamefont{{Bock}}},
  \bibinfo{author}{\bibfnamefont{J.~R.} \bibnamefont{{Bond}}},
  \bibinfo{author}{\bibfnamefont{J.}~\bibnamefont{{Brevik}}},
  \bibinfo{author}{\bibfnamefont{S.}~\bibnamefont{{Bryan}}},
  \bibinfo{author}{\bibfnamefont{C.~R.} \bibnamefont{{Contaldi}}},
  \bibnamefont{et~al.} (\bibinfo{year}{2007}), \eprint{arXiv:0710.0375}.

\bibitem[{\citenamefont{{Ruhl} et~al.}(2004)\citenamefont{{Ruhl}, {Ade},
  {Carlstrom}, {Cho}, {Crawford}, {Dobbs}, {Greer}, {Halverson}, {Holzapfel},
  {Lanting} et~al.}}]{Ruhl:2004kv}
\bibinfo{author}{\bibfnamefont{J.}~\bibnamefont{{Ruhl}}},
  \bibinfo{author}{\bibfnamefont{P.~A.~R.} \bibnamefont{{Ade}}},
  \bibinfo{author}{\bibfnamefont{J.~E.} \bibnamefont{{Carlstrom}}},
  \bibinfo{author}{\bibfnamefont{H.-M.} \bibnamefont{{Cho}}},
  \bibinfo{author}{\bibfnamefont{T.}~\bibnamefont{{Crawford}}},
  \bibinfo{author}{\bibfnamefont{M.}~\bibnamefont{{Dobbs}}},
  \bibinfo{author}{\bibfnamefont{C.~H.} \bibnamefont{{Greer}}},
  \bibinfo{author}{\bibfnamefont{N.~w.} \bibnamefont{{Halverson}}},
  \bibinfo{author}{\bibfnamefont{W.~L.} \bibnamefont{{Holzapfel}}},
  \bibinfo{author}{\bibfnamefont{T.~M.} \bibnamefont{{Lanting}}},
  \bibnamefont{et~al.}, in \emph{\bibinfo{booktitle}{Millimeter and
  Submillimeter Detectors for Astronomy II, Proceedings of the SPIE}}, edited
  by \bibinfo{editor}{\bibfnamefont{C.~M.} \bibnamefont{{Bradford}}},
  \bibinfo{editor}{\bibfnamefont{P.~A.~R.} \bibnamefont{{Ade}}},
  \bibinfo{editor}{\bibfnamefont{J.~E.} \bibnamefont{{Aguirre}}},
  \bibinfo{editor}{\bibfnamefont{J.~J.} \bibnamefont{{Bock}}},
  \bibinfo{editor}{\bibfnamefont{M.}~\bibnamefont{{Dragovan}}},
  \bibinfo{editor}{\bibfnamefont{L.}~\bibnamefont{{Duband}}},
  \bibinfo{editor}{\bibfnamefont{L.}~\bibnamefont{{Earle}}},
  \bibinfo{editor}{\bibfnamefont{J.}~\bibnamefont{{Glenn}}},
  \bibinfo{editor}{\bibfnamefont{H.}~\bibnamefont{{Matsuhara}}},
  \bibinfo{editor}{\bibfnamefont{B.~J.} \bibnamefont{{Naylor}}},
  \bibnamefont{et~al.} (\bibinfo{year}{2004}), vol. \bibinfo{volume}{5498}, pp.
  \bibinfo{pages}{11--29}, \eprint{arXiv:astro-ph/0411122}.

\bibitem[{\citenamefont{Kogut et~al.}(2006)}]{Kogut:2006nd}
\bibinfo{author}{\bibfnamefont{A.}~\bibnamefont{Kogut}} \bibnamefont{et~al.},
  \bibinfo{journal}{New Astron. Rev.} \textbf{\bibinfo{volume}{50}},
  \bibinfo{pages}{1009} (\bibinfo{year}{2006}),
  \eprint{arXiv:astro-ph/0609546}.

\bibitem[{\citenamefont{{Polenta} et~al.}(2007)\citenamefont{{Polenta}, {Ade},
  {Bartlett}, {Br{\'e}elle}, {Conversi}, {de Bernardis}, {Dufour}, {Gervasi},
  {Giard}, {Giordano} et~al.}}]{Polenta07}
\bibinfo{author}{\bibfnamefont{G.}~\bibnamefont{{Polenta}}},
  \bibinfo{author}{\bibfnamefont{P.~A.~R.} \bibnamefont{{Ade}}},
  \bibinfo{author}{\bibfnamefont{J.}~\bibnamefont{{Bartlett}}},
  \bibinfo{author}{\bibfnamefont{E.}~\bibnamefont{{Br{\'e}elle}}},
  \bibinfo{author}{\bibfnamefont{L.}~\bibnamefont{{Conversi}}},
  \bibinfo{author}{\bibfnamefont{P.}~\bibnamefont{{de Bernardis}}},
  \bibinfo{author}{\bibfnamefont{C.}~\bibnamefont{{Dufour}}},
  \bibinfo{author}{\bibfnamefont{M.}~\bibnamefont{{Gervasi}}},
  \bibinfo{author}{\bibfnamefont{M.}~\bibnamefont{{Giard}}},
  \bibinfo{author}{\bibfnamefont{C.}~\bibnamefont{{Giordano}}},
  \bibnamefont{et~al.}, \bibinfo{journal}{New Astronomy Review}
  \textbf{\bibinfo{volume}{51}}, \bibinfo{pages}{256} (\bibinfo{year}{2007}).

\bibitem[{\citenamefont{{Zaldarriaga}}(1997)}]{Zal97}
\bibinfo{author}{\bibfnamefont{M.}~\bibnamefont{{Zaldarriaga}}},
  \bibinfo{journal}{\prd} \textbf{\bibinfo{volume}{55}}, \bibinfo{pages}{1822}
  (\bibinfo{year}{1997}), \eprint{arXiv:astro-ph/9608050}.

\bibitem[{\citenamefont{{Page} et~al.}(2007)\citenamefont{{Page}, {Hinshaw},
  {Komatsu}, {Nolta}, {Spergel}, {Bennett}, {Barnes}, {Bean}, {Dor{\'e}},
  {Dunkley} et~al.}}]{Pagetal07}
\bibinfo{author}{\bibfnamefont{L.}~\bibnamefont{{Page}}},
  \bibinfo{author}{\bibfnamefont{G.}~\bibnamefont{{Hinshaw}}},
  \bibinfo{author}{\bibfnamefont{E.}~\bibnamefont{{Komatsu}}},
  \bibinfo{author}{\bibfnamefont{M.~R.} \bibnamefont{{Nolta}}},
  \bibinfo{author}{\bibfnamefont{D.~N.} \bibnamefont{{Spergel}}},
  \bibinfo{author}{\bibfnamefont{C.~L.} \bibnamefont{{Bennett}}},
  \bibinfo{author}{\bibfnamefont{C.}~\bibnamefont{{Barnes}}},
  \bibinfo{author}{\bibfnamefont{R.}~\bibnamefont{{Bean}}},
  \bibinfo{author}{\bibfnamefont{O.}~\bibnamefont{{Dor{\'e}}}},
  \bibinfo{author}{\bibfnamefont{J.}~\bibnamefont{{Dunkley}}},
  \bibnamefont{et~al.}, \bibinfo{journal}{\apjs}
  \textbf{\bibinfo{volume}{170}}, \bibinfo{pages}{335} (\bibinfo{year}{2007}),
  \eprint{arXiv:astro-ph/0603450}.

\bibitem[{\citenamefont{{Holder} et~al.}(2003)\citenamefont{{Holder}, {Haiman},
  {Kaplinghat}, and {Knox}}}]{Holetal03}
\bibinfo{author}{\bibfnamefont{G.~P.} \bibnamefont{{Holder}}},
  \bibinfo{author}{\bibfnamefont{Z.}~\bibnamefont{{Haiman}}},
  \bibinfo{author}{\bibfnamefont{M.}~\bibnamefont{{Kaplinghat}}},
  \bibnamefont{and} \bibinfo{author}{\bibfnamefont{L.}~\bibnamefont{{Knox}}},
  \bibinfo{journal}{\apj} \textbf{\bibinfo{volume}{595}}, \bibinfo{pages}{13}
  (\bibinfo{year}{2003}), \eprint{arXiv:astro-ph/0302404}.

\bibitem[{\citenamefont{{Mortonson} and {Hu}}(2007)}]{MorHu07b}
\bibinfo{author}{\bibfnamefont{M.~J.} \bibnamefont{{Mortonson}}}
  \bibnamefont{and} \bibinfo{author}{\bibfnamefont{W.}~\bibnamefont{{Hu}}},
  \bibinfo{journal}{\apj (in press)}  (\bibinfo{year}{2007}),
  \eprint{arXiv:0705.1132}.

\bibitem[{\citenamefont{{Keating} and {Miller}}(2006)}]{KeaMil06}
\bibinfo{author}{\bibfnamefont{B.}~\bibnamefont{{Keating}}} \bibnamefont{and}
  \bibinfo{author}{\bibfnamefont{N.}~\bibnamefont{{Miller}}},
  \bibinfo{journal}{New Astronomy Review} \textbf{\bibinfo{volume}{50}},
  \bibinfo{pages}{184} (\bibinfo{year}{2006}), \eprint{arXiv:astro-ph/0508269}.

\bibitem[{\citenamefont{{Hu} and {Holder}}(2003)}]{HuHol03}
\bibinfo{author}{\bibfnamefont{W.}~\bibnamefont{{Hu}}} \bibnamefont{and}
  \bibinfo{author}{\bibfnamefont{G.~P.} \bibnamefont{{Holder}}},
  \bibinfo{journal}{\prd} \textbf{\bibinfo{volume}{68}},
  \bibinfo{pages}{023001} (\bibinfo{year}{2003}),
  \eprint{arXiv:astro-ph/0303400}.

\bibitem[{\citenamefont{{Lidsey} et~al.}(1997)\citenamefont{{Lidsey}, {Liddle},
  {Kolb}, {Copeland}, {Barreiro}, and {Abney}}}]{Lidetal97}
\bibinfo{author}{\bibfnamefont{J.~E.} \bibnamefont{{Lidsey}}},
  \bibinfo{author}{\bibfnamefont{A.~R.} \bibnamefont{{Liddle}}},
  \bibinfo{author}{\bibfnamefont{E.~W.} \bibnamefont{{Kolb}}},
  \bibinfo{author}{\bibfnamefont{E.~J.} \bibnamefont{{Copeland}}},
  \bibinfo{author}{\bibfnamefont{T.}~\bibnamefont{{Barreiro}}},
  \bibnamefont{and} \bibinfo{author}{\bibfnamefont{M.}~\bibnamefont{{Abney}}},
  \bibinfo{journal}{Reviews of Modern Physics} \textbf{\bibinfo{volume}{69}},
  \bibinfo{pages}{373} (\bibinfo{year}{1997}), \eprint{arXiv:astro-ph/9508078}.

\bibitem[{\citenamefont{{Zaldarriaga} and {Seljak}}(1998)}]{ZalSel98}
\bibinfo{author}{\bibfnamefont{M.}~\bibnamefont{{Zaldarriaga}}}
  \bibnamefont{and} \bibinfo{author}{\bibfnamefont{U.}~\bibnamefont{{Seljak}}},
  \bibinfo{journal}{\prd} \textbf{\bibinfo{volume}{58}},
  \bibinfo{pages}{023003} (\bibinfo{year}{1998}),
  \eprint{arXiv:astro-ph/9803150}.

\bibitem[{\citenamefont{{Bowden} et~al.}(2004)\citenamefont{{Bowden}, {Taylor},
  {Ganga}, {Ade}, {Bock}, {Cahill}, {Carlstrom}, {Church}, {Gear}, {Hinderks}
  et~al.}}]{Bowetal04}
\bibinfo{author}{\bibfnamefont{M.}~\bibnamefont{{Bowden}}},
  \bibinfo{author}{\bibfnamefont{A.~N.} \bibnamefont{{Taylor}}},
  \bibinfo{author}{\bibfnamefont{K.~M.} \bibnamefont{{Ganga}}},
  \bibinfo{author}{\bibfnamefont{P.~A.~R.} \bibnamefont{{Ade}}},
  \bibinfo{author}{\bibfnamefont{J.~J.} \bibnamefont{{Bock}}},
  \bibinfo{author}{\bibfnamefont{G.}~\bibnamefont{{Cahill}}},
  \bibinfo{author}{\bibfnamefont{J.~E.} \bibnamefont{{Carlstrom}}},
  \bibinfo{author}{\bibfnamefont{S.~E.} \bibnamefont{{Church}}},
  \bibinfo{author}{\bibfnamefont{W.~K.} \bibnamefont{{Gear}}},
  \bibinfo{author}{\bibfnamefont{J.~R.} \bibnamefont{{Hinderks}}},
  \bibnamefont{et~al.}, \bibinfo{journal}{\mnras}
  \textbf{\bibinfo{volume}{349}}, \bibinfo{pages}{321} (\bibinfo{year}{2004}),
  \eprint{arXiv:astro-ph/0309610}.

\bibitem[{\citenamefont{{Verde} et~al.}(2006)\citenamefont{{Verde}, {Peiris},
  and {Jimenez}}}]{VerPeiJim06}
\bibinfo{author}{\bibfnamefont{L.}~\bibnamefont{{Verde}}},
  \bibinfo{author}{\bibfnamefont{H.~V.} \bibnamefont{{Peiris}}},
  \bibnamefont{and}
  \bibinfo{author}{\bibfnamefont{R.}~\bibnamefont{{Jimenez}}},
  \bibinfo{journal}{Journal of Cosmology and Astro-Particle Physics}
  \textbf{\bibinfo{volume}{1}}, \bibinfo{pages}{19} (\bibinfo{year}{2006}),
  \eprint{arXiv:astro-ph/0506036}.

\bibitem[{\citenamefont{{Tucci} et~al.}(2005)\citenamefont{{Tucci},
  {Mart{\'{\i}}nez-Gonz{\'a}lez}, {Vielva}, and {Delabrouille}}}]{Tucetal05}
\bibinfo{author}{\bibfnamefont{M.}~\bibnamefont{{Tucci}}},
  \bibinfo{author}{\bibfnamefont{E.}~\bibnamefont{{Mart{\'{\i}}nez-Gonz{\'a}le%
z}}}, \bibinfo{author}{\bibfnamefont{P.}~\bibnamefont{{Vielva}}},
  \bibnamefont{and}
  \bibinfo{author}{\bibfnamefont{J.}~\bibnamefont{{Delabrouille}}},
  \bibinfo{journal}{\mnras} \textbf{\bibinfo{volume}{360}},
  \bibinfo{pages}{935} (\bibinfo{year}{2005}), \eprint{arXiv:astro-ph/0411567}.

\bibitem[{\citenamefont{{Amblard} et~al.}(2007)\citenamefont{{Amblard},
  {Cooray}, and {Kaplinghat}}}]{AmbCooKap07}
\bibinfo{author}{\bibfnamefont{A.}~\bibnamefont{{Amblard}}},
  \bibinfo{author}{\bibfnamefont{A.}~\bibnamefont{{Cooray}}}, \bibnamefont{and}
  \bibinfo{author}{\bibfnamefont{M.}~\bibnamefont{{Kaplinghat}}},
  \bibinfo{journal}{\prd} \textbf{\bibinfo{volume}{75}},
  \bibinfo{pages}{083508} (\bibinfo{year}{2007}),
  \eprint{arXiv:astro-ph/0610829}.

\bibitem[{\citenamefont{{Hu} and {Okamoto}}(2002)}]{HuOka01}
\bibinfo{author}{\bibfnamefont{W.}~\bibnamefont{{Hu}}} \bibnamefont{and}
  \bibinfo{author}{\bibfnamefont{T.}~\bibnamefont{{Okamoto}}},
  \bibinfo{journal}{\apj} \textbf{\bibinfo{volume}{574}}, \bibinfo{pages}{566}
  (\bibinfo{year}{2002}), \eprint{arXiv:astro-ph/0111606}.

\bibitem[{\citenamefont{Knox and Song}(2002)}]{KnoSon02}
\bibinfo{author}{\bibfnamefont{L.}~\bibnamefont{Knox}} \bibnamefont{and}
  \bibinfo{author}{\bibfnamefont{Y.-S.} \bibnamefont{Song}},
  \bibinfo{journal}{Phys. Rev. Lett.} \textbf{\bibinfo{volume}{89}},
  \bibinfo{pages}{011303} (\bibinfo{year}{2002}),
  \eprint{arXiv:astro-ph/0202286}.

\bibitem[{\citenamefont{{Kesden} et~al.}(2002)\citenamefont{{Kesden}, {Cooray},
  and {Kamionkowski}}}]{KesCooKam02}
\bibinfo{author}{\bibfnamefont{M.}~\bibnamefont{{Kesden}}},
  \bibinfo{author}{\bibfnamefont{A.}~\bibnamefont{{Cooray}}}, \bibnamefont{and}
  \bibinfo{author}{\bibfnamefont{M.}~\bibnamefont{{Kamionkowski}}},
  \bibinfo{journal}{\prl} \textbf{\bibinfo{volume}{89}}, \bibinfo{pages}{1304}
  (\bibinfo{year}{2002}).

\bibitem[{\citenamefont{Seljak and Hirata}(2004)}]{SelHir03}
\bibinfo{author}{\bibfnamefont{U.}~\bibnamefont{Seljak}} \bibnamefont{and}
  \bibinfo{author}{\bibfnamefont{C.~M.} \bibnamefont{Hirata}},
  \bibinfo{journal}{Phys. Rev.} \textbf{\bibinfo{volume}{D69}},
  \bibinfo{pages}{043005} (\bibinfo{year}{2004}),
  \eprint{arXiv:astro-ph/0310163}.

\bibitem[{\citenamefont{Marian and Bernstein}(2007)}]{Marian:2007sr}
\bibinfo{author}{\bibfnamefont{L.}~\bibnamefont{Marian}} \bibnamefont{and}
  \bibinfo{author}{\bibfnamefont{G.~M.} \bibnamefont{Bernstein}}
  (\bibinfo{year}{2007}), \eprint{arXiv:0710.2538}.

\bibitem[{\citenamefont{{Kaplinghat} et~al.}(2003)\citenamefont{{Kaplinghat},
  {Chu}, {Haiman}, {Holder}, {Knox}, and {Skordis}}}]{Kapetal03}
\bibinfo{author}{\bibfnamefont{M.}~\bibnamefont{{Kaplinghat}}},
  \bibinfo{author}{\bibfnamefont{M.}~\bibnamefont{{Chu}}},
  \bibinfo{author}{\bibfnamefont{Z.}~\bibnamefont{{Haiman}}},
  \bibinfo{author}{\bibfnamefont{G.~P.} \bibnamefont{{Holder}}},
  \bibinfo{author}{\bibfnamefont{L.}~\bibnamefont{{Knox}}}, \bibnamefont{and}
  \bibinfo{author}{\bibfnamefont{C.}~\bibnamefont{{Skordis}}},
  \bibinfo{journal}{\apj} \textbf{\bibinfo{volume}{583}}, \bibinfo{pages}{24}
  (\bibinfo{year}{2003}), \eprint{arXiv:astro-ph/0207591}.

\bibitem[{\citenamefont{{Colombo} et~al.}(2005)\citenamefont{{Colombo},
  {Bernardi}, {Casarini}, {Mainini}, {Bonometto}, {Carretti}, and
  {Fabbri}}}]{Coletal05}
\bibinfo{author}{\bibfnamefont{L.~P.~L.} \bibnamefont{{Colombo}}},
  \bibinfo{author}{\bibfnamefont{G.}~\bibnamefont{{Bernardi}}},
  \bibinfo{author}{\bibfnamefont{L.}~\bibnamefont{{Casarini}}},
  \bibinfo{author}{\bibfnamefont{R.}~\bibnamefont{{Mainini}}},
  \bibinfo{author}{\bibfnamefont{S.~A.} \bibnamefont{{Bonometto}}},
  \bibinfo{author}{\bibfnamefont{E.}~\bibnamefont{{Carretti}}},
  \bibnamefont{and} \bibinfo{author}{\bibfnamefont{R.}~\bibnamefont{{Fabbri}}},
  \bibinfo{journal}{\aap} \textbf{\bibinfo{volume}{435}}, \bibinfo{pages}{413}
  (\bibinfo{year}{2005}), \eprint{arXiv:astro-ph/0408022}.

\bibitem[{\citenamefont{{Hu} and {White}}(1997)}]{HuWhi97c}
\bibinfo{author}{\bibfnamefont{W.}~\bibnamefont{{Hu}}} \bibnamefont{and}
  \bibinfo{author}{\bibfnamefont{M.}~\bibnamefont{{White}}},
  \bibinfo{journal}{\prd} \textbf{\bibinfo{volume}{56}}, \bibinfo{pages}{596}
  (\bibinfo{year}{1997}), \eprint{arXiv:astro-ph/9702170}.

\bibitem[{\citenamefont{Christensen et~al.}(2001)\citenamefont{Christensen,
  Meyer, Knox, and Luey}}]{Chretal01}
\bibinfo{author}{\bibfnamefont{N.}~\bibnamefont{Christensen}},
  \bibinfo{author}{\bibfnamefont{R.}~\bibnamefont{Meyer}},
  \bibinfo{author}{\bibfnamefont{L.}~\bibnamefont{Knox}}, \bibnamefont{and}
  \bibinfo{author}{\bibfnamefont{B.}~\bibnamefont{Luey}},
  \bibinfo{journal}{Class. Quant. Grav.} \textbf{\bibinfo{volume}{18}},
  \bibinfo{pages}{2677} (\bibinfo{year}{2001}),
  \eprint{arXiv:astro-ph/0103134}.

\bibitem[{\citenamefont{Kosowsky et~al.}(2002)\citenamefont{Kosowsky,
  Milosavljevic, and Jimenez}}]{KosMilJim02}
\bibinfo{author}{\bibfnamefont{A.}~\bibnamefont{Kosowsky}},
  \bibinfo{author}{\bibfnamefont{M.}~\bibnamefont{Milosavljevic}},
  \bibnamefont{and} \bibinfo{author}{\bibfnamefont{R.}~\bibnamefont{Jimenez}},
  \bibinfo{journal}{Phys. Rev.} \textbf{\bibinfo{volume}{D66}},
  \bibinfo{pages}{063007} (\bibinfo{year}{2002}),
  \eprint{arXiv:astro-ph/0206014}.

\bibitem[{\citenamefont{Dunkley et~al.}(2005)\citenamefont{Dunkley, Bucher,
  Ferreira, Moodley, and Skordis}}]{Dunetal04}
\bibinfo{author}{\bibfnamefont{J.}~\bibnamefont{Dunkley}},
  \bibinfo{author}{\bibfnamefont{M.}~\bibnamefont{Bucher}},
  \bibinfo{author}{\bibfnamefont{P.~G.} \bibnamefont{Ferreira}},
  \bibinfo{author}{\bibfnamefont{K.}~\bibnamefont{Moodley}}, \bibnamefont{and}
  \bibinfo{author}{\bibfnamefont{C.}~\bibnamefont{Skordis}},
  \bibinfo{journal}{Mon. Not. Roy. Astron. Soc.}
  \textbf{\bibinfo{volume}{356}}, \bibinfo{pages}{925} (\bibinfo{year}{2005}),
  \eprint{arXiv:astro-ph/0405462}.

\bibitem[{\citenamefont{{Lewis} and {Bridle}}(2002)}]{LewBri02}
\bibinfo{author}{\bibfnamefont{A.}~\bibnamefont{{Lewis}}} \bibnamefont{and}
  \bibinfo{author}{\bibfnamefont{S.}~\bibnamefont{{Bridle}}},
  \bibinfo{journal}{\prd} \textbf{\bibinfo{volume}{66}},
  \bibinfo{pages}{103511} (\bibinfo{year}{2002}),
  \eprint{arXiv:astro-ph/0205436}.

\bibitem[{\citenamefont{{Lewis} et~al.}(2000)\citenamefont{{Lewis},
  {Challinor}, and {Lasenby}}}]{Lewetal00}
\bibinfo{author}{\bibfnamefont{A.}~\bibnamefont{{Lewis}}},
  \bibinfo{author}{\bibfnamefont{A.}~\bibnamefont{{Challinor}}},
  \bibnamefont{and}
  \bibinfo{author}{\bibfnamefont{A.}~\bibnamefont{{Lasenby}}},
  \bibinfo{journal}{\apj} \textbf{\bibinfo{volume}{538}}, \bibinfo{pages}{473}
  (\bibinfo{year}{2000}), \eprint{arXiv:astro-ph/9911177}.

\bibitem[{\citenamefont{{Gelman} and {Rubin}}(1992)}]{GelRub92}
\bibinfo{author}{\bibfnamefont{A.}~\bibnamefont{{Gelman}}} \bibnamefont{and}
  \bibinfo{author}{\bibfnamefont{D.~B.} \bibnamefont{{Rubin}}},
  \bibinfo{journal}{Statistical Science} \textbf{\bibinfo{volume}{7}},
  \bibinfo{pages}{457} (\bibinfo{year}{1992}), ISSN \bibinfo{issn}{0883-4237}.

\bibitem[{\citenamefont{{Brooks} and {Gelman}}(1998)}]{BroGel98}
\bibinfo{author}{\bibfnamefont{S.~P.} \bibnamefont{{Brooks}}} \bibnamefont{and}
  \bibinfo{author}{\bibfnamefont{A.}~\bibnamefont{{Gelman}}},
  \bibinfo{journal}{Journal of Computational and Graphical Statistics}
  \textbf{\bibinfo{volume}{7}}, \bibinfo{pages}{434} (\bibinfo{year}{1998}),
  ISSN \bibinfo{issn}{1061-8600}.

\bibitem[{\citenamefont{Raftery and Lewis}(1992)}]{RafLew92}
\bibinfo{author}{\bibfnamefont{A.~E.} \bibnamefont{Raftery}} \bibnamefont{and}
  \bibinfo{author}{\bibfnamefont{S.~M.} \bibnamefont{Lewis}}, in
  \emph{\bibinfo{booktitle}{Bayesian Statistics}}, edited by
  \bibinfo{editor}{\bibfnamefont{J.~M.} \bibnamefont{Bernado}}
  (\bibinfo{publisher}{OUP}, \bibinfo{year}{1992}), p. \bibinfo{pages}{765}.

\bibitem[{\citenamefont{{Albrecht} et~al.}(2006)\citenamefont{{Albrecht},
  {Bernstein}, {Cahn}, {Freedman}, {Hewitt}, {Hu}, {Huth}, {Kamionkowski},
  {Kolb}, {Knox} et~al.}}]{Albetal06}
\bibinfo{author}{\bibfnamefont{A.}~\bibnamefont{{Albrecht}}},
  \bibinfo{author}{\bibfnamefont{G.}~\bibnamefont{{Bernstein}}},
  \bibinfo{author}{\bibfnamefont{R.}~\bibnamefont{{Cahn}}},
  \bibinfo{author}{\bibfnamefont{W.~L.} \bibnamefont{{Freedman}}},
  \bibinfo{author}{\bibfnamefont{J.}~\bibnamefont{{Hewitt}}},
  \bibinfo{author}{\bibfnamefont{W.}~\bibnamefont{{Hu}}},
  \bibinfo{author}{\bibfnamefont{J.}~\bibnamefont{{Huth}}},
  \bibinfo{author}{\bibfnamefont{M.}~\bibnamefont{{Kamionkowski}}},
  \bibinfo{author}{\bibfnamefont{E.~W.} \bibnamefont{{Kolb}}},
  \bibinfo{author}{\bibfnamefont{L.}~\bibnamefont{{Knox}}},
  \bibnamefont{et~al.} (\bibinfo{year}{2006}), \eprint{arXiv:astro-ph/0609591}.

\bibitem[{\citenamefont{{Smith}
  et~al.}(2006{\natexlab{a}})\citenamefont{{Smith}, {Hu}, and
  {Kaplinghat}}}]{SmiHuKap06}
\bibinfo{author}{\bibfnamefont{K.~M.} \bibnamefont{{Smith}}},
  \bibinfo{author}{\bibfnamefont{W.}~\bibnamefont{{Hu}}}, \bibnamefont{and}
  \bibinfo{author}{\bibfnamefont{M.}~\bibnamefont{{Kaplinghat}}},
  \bibinfo{journal}{\prd} \textbf{\bibinfo{volume}{74}},
  \bibinfo{pages}{123002} (\bibinfo{year}{2006}{\natexlab{a}}),
  \eprint{arXiv:astro-ph/0607315}.

\bibitem[{\citenamefont{{Lewis} et~al.}(2006)\citenamefont{{Lewis}, {Weller},
  and {Battye}}}]{LewWelBat06}
\bibinfo{author}{\bibfnamefont{A.}~\bibnamefont{{Lewis}}},
  \bibinfo{author}{\bibfnamefont{J.}~\bibnamefont{{Weller}}}, \bibnamefont{and}
  \bibinfo{author}{\bibfnamefont{R.}~\bibnamefont{{Battye}}},
  \bibinfo{journal}{\mnras} \textbf{\bibinfo{volume}{373}},
  \bibinfo{pages}{561} (\bibinfo{year}{2006}), \eprint{arXiv:astro-ph/0606552}.

\bibitem[{\citenamefont{{Chongchitnan} and {Efstathiou}}(2006)}]{ChoEfs06}
\bibinfo{author}{\bibfnamefont{S.}~\bibnamefont{{Chongchitnan}}}
  \bibnamefont{and}
  \bibinfo{author}{\bibfnamefont{G.}~\bibnamefont{{Efstathiou}}},
  \bibinfo{journal}{\prd} \textbf{\bibinfo{volume}{73}},
  \bibinfo{pages}{083511} (\bibinfo{year}{2006}),
  \eprint{arXiv:astro-ph/0602594}.

\bibitem[{\citenamefont{{Smith}
  et~al.}(2006{\natexlab{b}})\citenamefont{{Smith}, {Peiris}, and
  {Cooray}}}]{SmiPeiCoo06}
\bibinfo{author}{\bibfnamefont{T.~L.} \bibnamefont{{Smith}}},
  \bibinfo{author}{\bibfnamefont{H.~V.} \bibnamefont{{Peiris}}},
  \bibnamefont{and} \bibinfo{author}{\bibfnamefont{A.}~\bibnamefont{{Cooray}}},
  \bibinfo{journal}{\prd} \textbf{\bibinfo{volume}{73}},
  \bibinfo{pages}{123503} (\bibinfo{year}{2006}{\natexlab{b}}),
  \eprint{arXiv:astro-ph/0602137}.

\end{thebibliography}

\end{document}